\documentclass[prd,floats,preprint,superscriptaddress,eqsecnum,floatfix,nofootinbib,12pt]{revtex4}
\pdfoutput=1
\usepackage{amsmath}
\usepackage{graphics, graphicx}
\usepackage{epsfig}
\usepackage{ulem}
\usepackage{color}

\renewcommand{\Re}{{\rm Re}}

\def\1p{{(1p)}}
\def\be{\begin{equation}}
\def\ee{\end{equation}}
\def\beq{\begin{eqnarray}}
\def\eeq{\end{eqnarray}}
\def\cf{}

\def\p0{\phi_0}
\def\z0{\zeta_0}

\def\epb{{(2)}}

\def\uf{}

\def\vx{{\vec x}}
\def\lm{\lambda_{-}}
\def\lp{\lambda_{+}}
\def\th{{\tilde h}}

\def\cf{}

\newcommand{\ttle}[1]{{\it #1}}

\begin{document}

\vspace{1cm}

\title{Vector Fields in Holographic Cosmology}

\author{James B.  Hartle}
\affiliation{Department of Physics, University of California, Santa Barbara,  93106, USA}
\author{S.W. Hawking}
\affiliation{DAMTP, CMS, Wilberforce Road, CB3 0WA Cambridge, UK}
\author{Thomas Hertog}
\affiliation{Institute for Theoretical Physics, KU Leuven, 3001 Leuven, Belgium}

\bibliographystyle{unsrt}

\vspace{1cm}

\begin{abstract}

We extend the holographic formulation of the semiclassical no-boundary wave function (NBWF) to models with Maxwell vector fields. It is shown that the familiar saddle points of the NBWF have a representation in which a regular, Euclidean asymptotic AdS geometry smoothly joins onto a Lorentzian asymptotically de Sitter universe through a complex transition region. The tree level probabilities of Lorentzian histories are fully specified by the action of the AdS region of the saddle points. The scalar and vector matter profiles in this region are complex from an AdS viewpoint, with universal asymptotic phases. The dual description of the semiclassical NBWF thus involves complex deformations of Euclidean CFTs. 

\end{abstract}

\vskip.8in
\vspace{1cm}


\maketitle

\section{Introduction}
\label{intro}

In cosmology one is interested in computing the probability measure for different configurations of geometry and fields on a spacelike surface $\Sigma$. For a given dynamical model this measure is given by the universe's quantum state \cite{HH83}. In a series of papers  \cite{HHH08,HHH08b} we have calculated the tree level measure predicted by the no-boundary wave function (NBWF) for gravity coupled to a positive cosmological constant $\Lambda$ and a scalar field with a positive potential. Predictions for our observations are obtained by further conditioning on our observational situation and its possible locations in each history, and then summing over what is unobserved \cite{HHH10}. 

The predictions of the semiclassical NBWF are in good agreement with recent observations in certain landscape models that contain regions of eternal inflation \cite{H13}. It is therefore of interest to find a mathematically more precise formulation of the NBWF that allows one to reliably calculate the probability measure beyond the saddle point approximation. To this end we have recently developed a holographic formulation of the NBWF \cite{HH11}, for gravity coupled to a cosmological constant and a scalar field. Here we generalize this to matter with Maxwell vector fields. 

The development of this holographic framework for quantum cosmology fits in a broader program on holographic cosmology that aims to use the {\uf general} insights that have emerged from AdS/CFT to place cosmology on firm theoretical footing\footnote{Previous studies of the connection between Euclidean AdS and de Sitter from a wave function of the universe perspective include \cite{Bala02,Maldacena03,Fadden10,Harlow11,Maldacena11,Anninos12,Castro12,Anninos13}.}. A key feature of {\cf our} approach is that it does not involve a map of solutions from one theory to solutions of a different theory. Instead it makes use of the complex structure available in a given theory of the quantum state of the universe to establish a connection between the semiclassical NBWF in a cosmological setting and Euclidean AdS/CFT \cite{Witten98}. 

In its usual form the Euclidean AdS/CFT duality calculates the large volume limit of the wave function of the universe in a regime where it corresponds to Euclidean `histories' \cite{Horowitz04}. We recently showed \cite{HH11} that, in scalar field models, an extension of Euclidean AdS/CFT to complex saddle points yields the wave function in a domain where it describes a class of Lorentzian cosmologies.

Complex saddle points arise naturally in the semiclassical approximation to the NBWF where their action specifies the amplitude of different configurations. The saddle point action is an integral of its complex geometry and fields that includes an integral over time. Different complex contours for this time integral give different representations of the saddle point, each giving the same amplitude for the configuration it corresponds to. Using this freedom of choice of contour, we identified two different useful representations of the saddle points corresponding to Lorentzian, asymptotically de Sitter histories. In one representation (dS) the interior geometry behaves as though the cosmological constant and the scalar potential were positive. In the other (AdS) representation the Euclidean part of the interior geometry behaves as though $\Lambda$ and the potential are negative. The geometry in the AdS representation is that of a regular, AdS domain wall which joins smoothly onto the boundary configuration through a complex transition region in which the spatial part of the metric changes signature. 

The AdS representation of the familiar saddle points of the NBWF establishes a connection between the NBWF and the Euclidean AdS/CFT setup.
The relative probabilities of different boundary configurations are specified by the regularized action of the AdS region of the saddle points \cite{HH11}. Hence by applying Euclidean AdS/CFT, along the lines of \cite{Horowitz04}, one obtains a holographic form of the semiclassical NBWF in which the probabilities of asymptotically de Sitter histories are given by the partition function of (deformed) AdS/CFT dual Euclidean field theories. 

The scalar matter profile along the AdS domain wall regime of the saddle points is complex. This includes its asymptotic behavior in the AdS region, which enters as an external source in the dual partition function that turns on a deformation {\uf of the dual field theory action}. Hence the dual description of the no-boundary measure on classical configurations involves {\it complex deformations} of Euclidean CFTs familiar from AdS/CFT. The phases of the scalar sources in the dual are universal. {\uf That is, they }can be derived from a purely asymptotic analysis independently of the specific quantum state \cite{HH11}.

In this paper we extend our approach to matter models with vectors. In Section \ref{saddlept} we establish the AdS representation of general, inhomogeneous saddle points associated with asymptotically de Sitter histories in models with scalar and vector matter. We use this representation in Section \ref{saddlept2} to derive the probabilities of general asymptotically de Sitter histories from the regularized action of the AdS domain wall regime of the corresponding saddle points. The vector fields in the AdS region of the saddle points are purely imaginary from an AdS viewpoint. This implies that, in analogy with scalars, the AdS/CFT dual description of the NBWF in the presence of vectors involves complex deformations of Euclidean CFTs. We discuss the holographic form of the wave function in Section \ref{holo}. In Section \ref{homo} we illustrate our general framework with the wave function of vector perturbations in a homogeneous, expanding background which we calculate using both representations of the background saddle point. The NBWF predicts that the vector field remains in its ground state. In Section \ref{disc} we conclude with a few more general remarks on holographic cosmology.

\section{Different Representations of Complex Saddle Points}
\label{saddlept}

A quantum state of the universe is specified by a wave function $\Psi$ on the superspace of  three-geometries and matter field configurations on a spacelike boundary surface $\Sigma$ which we take to be closed. We consider matter models consisting of a single scalar field $\phi$ and a vector field $A_{\mu}$. Schematically we write $\Psi[h,\chi,\vec B]$ where $h$ stands for the metric $h_{ij}(\vec x)$ representing the boundary three-geometry, and $\chi(\vec x)$ and $\vec B(\vec x)$ are the boundary configurations\footnote{Hence in this paper $\vec B$ does not denote the magnetic field.} of $\phi$ and $A_{\mu}$. 
We assume the no-boundary wave function (NBWF) as a model of the quantum state \cite{HH83}. This is given by a sum over regular four-geometries $g_{\mu \nu}$ and matter fields $\phi$ and $A_{\mu}$ on a four-disk $M$ that match $(h,\chi,\vec B)$  on $\Sigma$. In the semiclassical approximation different  configurations {\uf labeled by} $(h,\chi,\vec B)$ are weighted by $\exp(-I/\hbar)$ where $I$ is the Euclidean action of a regular, complex extremal solution that matches the prescribed data on $\Sigma$.

A wave function of the universe predicts that configurations evolve classically when, with an appropriate coarse-graining, its phase varies rapidly compared to its amplitude \cite{HHH08}. To leading order in $\hbar$ both the phase and the amplitude are given by the action of the dominant saddle point. In this semiclassical approximation,
\begin{equation}
\Psi[h,\chi,\vec B] \approx  \exp\{(-I[h,\chi,\vec B] /\hbar\} = \exp\{(-I_R[h,\chi,\vec B] +i S[h,\chi,\vec B])/\hbar\} ,
\label{semiclass}
\end{equation}
where $I_R[h,\chi,\vec B]$ and $-S[h,\chi,\vec B]$ are the real and imaginary parts of the Euclidean action, evaluated at the saddle point. 
The probabilities of different classical configurations are proportional to $\exp(-2I_R/\hbar)$. They are conserved along a classical trajectory as a consequence of the Wheeler-DeWitt equation \cite{HHH08} and therefore yield a probability measure on the phase space of classical {\it histories}.

We  {\uf work in a four-dimensional bulk  and} take the Euclidean action $I[g(x),\phi(x),A(x)]$ to be a sum of the Einstein-Hilbert action $I_{EH}$ (in Planck units where $\hbar=c=G=1$)
\begin{equation}
I_{EH}[g] = -\frac{1}{16\pi}\int_M d^4 x (g)^{1/2}(R-2\Lambda) -\frac{1}{8\pi}\int_{\partial M} d^3x (h)^{1/2}K
\label{curvact}
\end{equation}
and the following matter action $I_M$
\begin{equation}
I_M [g,\phi,A]=\frac{1}{2} \int_M d^4x (g)^{1/2}[(\nabla\phi)^2 +2V(\phi)] +  \frac{1}{4} \int_M d^4x (g)^{1/2}F^{\mu \nu} F_{\mu \nu}
\label{mattact}
\end{equation}
where $F_{\mu \nu} = \partial_{\mu} A_{\nu} - \partial_{\nu} A_{\mu}$. The generalization to include a coupling between the scalar and vector fields does not affect our results so we neglect this here {\uf  for simplicity}.

We take the cosmological constant $\Lambda=3H^2$ and the scalar potential $V$ in the action \eqref{curvact}-\eqref{mattact} to be positive. This means that for positive signature metrics this is the Euclidean action of de Sitter gravity coupled to a scalar with a positive potential $V$ and a vector field $A_{\mu}$. For negative signature metrics this is minus the action of anti-de Sitter gravity coupled to a scalar with a negative potential $-V$ and a vector field $\tilde A_{\mu} \equiv iA_{\mu}$ that is imaginary from a de Sitter viewpoint\footnote{This relation under a signature reversal was noted long ago and used to establish a map from an AdS to a dS theory in \cite{Vilenkin88}, and more recently in \cite{Fadden10}.}. 

A global notion of signature is meaningless for the complex saddle point solutions that specify the semiclassical wave function \eqref{semiclass}. This is because the signature changes across the saddle point geometry. Moreover the signature in the saddle point interior differs depending on its representation. In scalar field models an analysis of the different saddle point representations led to a useful connection, described above, between (asymptotic) Lorentzian dS histories and Euclidean AdS geometries \cite{HH11}.
We now generalize this to models including vector matter, given by the action \eqref{curvact}-\eqref{mattact}.

The line element of a complex saddle point that specifies the amplitude of a general boundary configuration can be written as
\begin{equation}
ds^2=N^2(\lambda) d\lambda^2 +g_{ij}(\lambda,\vx) dx^i dx^j
\label{eucmetric_hij}
\end{equation} 
where $(\lambda, x^i)$ are four real coordinates on the real manifold $M$.  Conventionally we take $\lambda$ {\uf to be a `radial' coordinate so that}  $\lambda=0$ locates the origin or `South Pole' (SP) of the saddle point and take $\lambda=1$ to locate the boundary $\Sigma$ of $M$.  Saddle points may be represented by complex metrics -- complex $N$ and $g_{ij}$ -- but the coordinates  $(\lambda, x^i)$ are always real.  

The field equations derived from \eqref{curvact}-\eqref{mattact} can be solved for $g_{ij}(\lambda,\vx),\phi(\lambda,\vx)$ and $A_{\mu}(\lambda,\vx)$ for any complex $N(\lambda)$ that is specified. Different choices of $N(\lambda)$ therefore give different representations of the same saddle point. A convenient way to exhibit these different representations is to introduce the function $\tau(\lambda)$ defined by
\begin{equation}
\tau(\lambda) \equiv \int_0^\lambda d\lambda' N(\lambda').
\label{deftau}
\end{equation} 
Different choices of $N(\lambda)$ correspond to different contours in the complex $\tau$-plane. Contours start from the SP at $\lambda=\tau=0$ and end at the boundary $\lambda=1$ with $\tau(1)\equiv \upsilon$. Conversely, for any contour $\tau(\lambda)$ there is an $N(\lambda)\equiv d\tau(\lambda)/d\lambda$. Each contour connecting $\tau=0$ to $\tau=\upsilon$ is therefore a different representation of the same complex saddle point with the same action.

The semiclassical approximation to the wave function holds in the asymptotic `large volume' domain where the cosmological constant dominates the dynamics. Asymptotically, a semiclassical wave function specifies an ensemble of asymptotic solutions to the equations of motion. In terms of the complex time variable 
\be
u\equiv e^{iH\tau} = e^{-Hy+iHx} . 
\label{defu}
\ee
the asymptotic (small $u$) form of the general solution is given by
\begin{subequations}
\label{expansions}
\be
\label{hijexpn}
g_{ij}(u,\vx)=\frac{c^2}{u^2}[\th_{ij}(\vx) +\th_{ij}^{(2)}(\vx) u^2 + \th_{ij}^{(-)}(\vx)u^{2\lambda_{-}} +\th_{ij}^{(3)}(\vx)u^3 +\cdots] . 
\end{equation}
\be
\label{phia_hij}
\phi(u,\vx) = u^{\lm}(\alpha(\vx) +\alpha_1(\vx) u + \cdots)  +  u^{\lp}(\beta(\vx) +\beta_1(\vx) u +\cdots) . 
\ee
\be
\label{vectexp}
A_{i} (u,\vx)= B_{i} (\vx) + u C_{i}(\vx)+\cdots
\ee
\end{subequations}
where $\lambda_{\pm}\equiv \frac{3}{2}[1\pm \sqrt{1-(2m/3)^2}]$, and $\th_{ij}(\vx)$ is real and normalized to have unit volume thus determining the constant $c$.

The asymptotic solutions are locally determined from the asymptotic equations in terms of the arbitrary `boundary values' $c^2 \th_{ij}$, 
$\alpha (\vx)$ and $B_{i} (\vx)$, up to the $u^3$ term in \eqref{hijexpn}, to order $u^{\lambda_{+}^{\ }}$ in \eqref{phia_hij} and to leading order in \eqref{vectexp}. Beyond these terms in \eqref{expansions} the interior dynamics and the specific choice of quantum state become important.
 
\begin{figure}[t]
\includegraphics[width=1.7in]{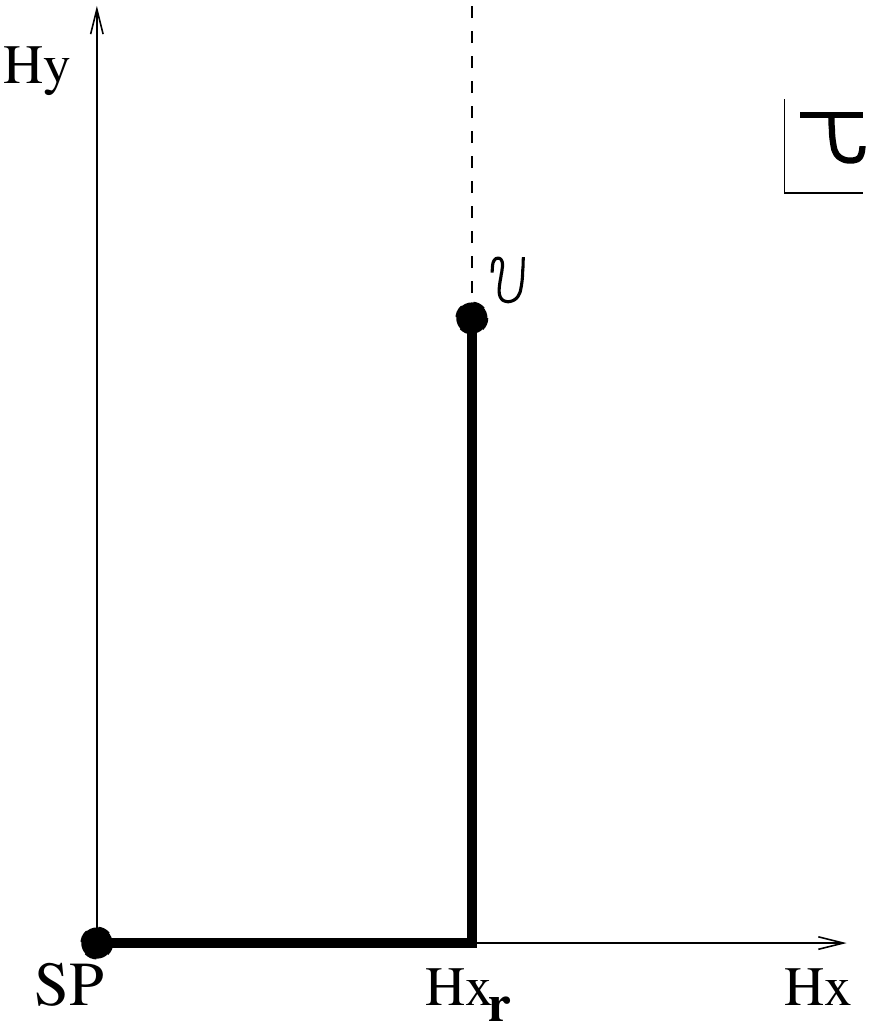}\hfill 
\includegraphics[width=1.8in]{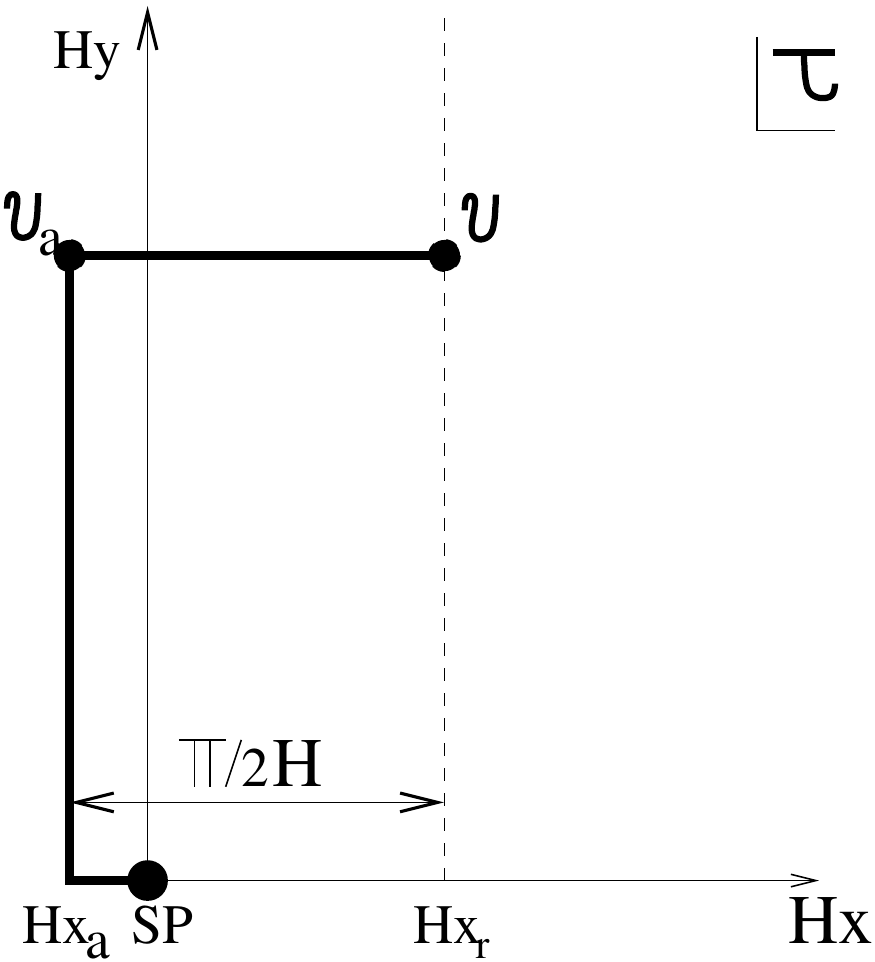}\hfill 
\includegraphics[width=1.2in]{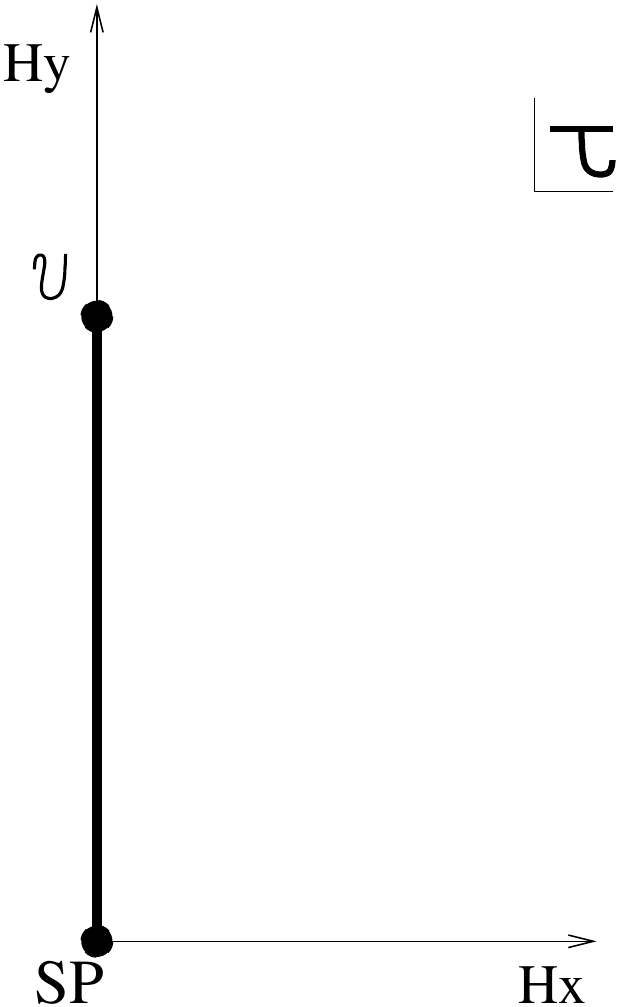}  
\caption{{\it Left panel}: The contour $C_A$ in the complex $\tau$-plane of a de Sitter representation of a complex saddle point of the NBWF. Along the horizontal part the geometry is half a deformed Euclidean three-sphere. Along the vertical part it is asymptotically Lorentzian de Sitter space.\\ {\it Middle panel:} The contour $C_B$ that gives the AdS representation of the same complex saddle point. The vertical part is asymptotically Euclidean AdS with a complex matter profile. The horizontal part is a complex geometry that interpolates between asymptotic AdS and de Sitter. \\ {\it Right panel:} The contour $C_C$ that provides the standard AdS representation of real asymptotically Euclidean AdS saddle points that are not associated with classical histories.}  
\label{contours}
 \end{figure}

For a saddle point solution to contribute to the semiclassical NBWF {\cf the observables must take real values on $\Sigma$ \cite{HHH12,HHH12b}. Hence the metric $h$, the scalar $\chi$ and the frame fields -- the components of the vector field in an orthonormal frame -- must all be real at $\tau = \upsilon$. The frame fields are given by $\vec B/b$, where the `scale factor' $b \equiv c/u$, multiplied by a real matrix specified by $\tilde h$.

Reality of the observables on the boundary selects two classes of saddle points, }corresponding respectively to asymptotically Euclidean AdS and Lorentzian de Sitter histories \cite{HHH12}. The saddle points associated with de Sitter histories are found by tuning the phases of the fields so that $b$, $\phi$ and $A_{\mu}$ all tend to real functions along a vertical line $x=x_r$ in the complex $\tau$-plane \cite{HHH08,HH11}. Writing $c=|c|e^{i\theta_c}$, $\alpha = |\alpha|e^{i\theta_{\alpha}}$ and $B_i = |B|e^{i\theta_{B}}$ the expansions \eqref{hijexpn} - \eqref{vectexp} imply
\be
\label{phases}
\theta_c=x_r, \qquad \theta_{\alpha}=-\lambda_{-}^{\ }x_r , \qquad \theta_B = 0
\ee 

By choosing a particular contour connecting $\tau=0$ to $\tau=\upsilon = x_r + iy_{\upsilon}$ we obtain a concrete representation of the saddle point geometry. The contour $C_A$ in Fig \ref{contours}(a) provides an example. Along the horizontal part of this the geometry is the Euclidean geometry of half a deformed four-sphere. Along the vertical part from $(x_r, 0)$ to $(x_r, y_{\upsilon})$ the geometry tends to a Lorentzian space that is locally de Sitter,
\be
ds^2 \approx -dy^2 + \frac{1}{4H^2}e^{2Hy} \th_{ij}(\vx) dx^i dx^j.
\label{LordS}
\ee
Thus the contour $C_A$ gives the familiar representation of the no-boundary saddle points that are associated with asymptotically real, Lorentzian, de Sitter histories. The ensemble of histories of this kind predicted by the NBWF is known explicitly in a homogeneous and isotropic minisuperspace approximation with linear perturbations, where it consists of universes with an early period of scalar field driven inflation \cite{HHH08}. The saddle point action acquires a rapidly varying phase factor along the vertical part of the contour $C_A$ so the wave function takes a WKB form \eqref{semiclass} and predicts classical cosmological evolution.

There is, however, an alternative representation of the same class of saddle points given by the contour $C_B$ in Fig \ref{contours}(b). Along the $x=x_a =x_r-\pi/2H$ line one has
\be
\label{adsmetric}
ds^2 \approx -dy^2 - \frac{1}{4H^2}e^{2Hy} \th_{ij}(\vx) dx^i dx^j.
\ee
This is negative signature, Euclidean, asymptotically local AdS space. The asymptotic form of the matter fields along $x=x_a$ follows from \eqref{phia_hij} --\eqref{vectexp} and is given by
\be
\phi(y,\vx) \approx |\alpha (\vx)|e^{-i\lambda_{-}^{\ }\pi/2}e^{-\lambda_{-}^{\ }y} \equiv \tilde \alpha e^{-\lambda_{-}^{\ }y}, \qquad  A_{i} (y,\vx) \approx B_i (\vx) \equiv -i\tilde B_i
\label{phase}
\ee
where $\tilde B_i$ is the boundary value of $\tilde A_{\mu} \equiv iA_{\mu}$.
Therefore the condition that the scalar be real at the endpoint $\upsilon$, at $x=x_r$, means it is complex along the AdS branch of the contour, with an asymptotic phase given by \eqref{phase}. Similarly, the requirement that $A_{i}$ be real at the endpoint $\upsilon$ means that the usual AdS vector field $\tilde A_i = iA_i$ is purely imaginary in the asymptotic AdS region of de Sitter saddle points\footnote{\cf {Hence, as expected, Maxwell fields behave as conformally coupled scalars as far as their asymptotic phase is concerned.}}.

To summarize, in the `AdS representation' of Fig \ref{contours}(b) the saddle point geometry consists of a Euclidean, asymptotically locally AdS geometry with complex matter field profiles in the `radial' direction $y$. This is then joined smoothly onto the Lorentzian, asymptotically de Sitter geometry through a transition region -- corresponding to the horizontal part of the contour $C_B$ -- where the geometry is complex. The complex structure of the semiclassical NBWF thus provides a natural connection\footnote{The AdS/de Sitter connection exhibited here can be generalized to wave functions other than the NBWF, since it follows directly from the asymptotic structure of the wave function \cite{HHH12}. The connection therefore holds for any wave function satisfying the Wheeler-DeWitt equation.} between (asymptotically) Lorentzian de Sitter histories and Euclidean AdS geometries. 

As mentioned earlier there is a second set of saddle points {\cf with real observables on the boundary} \cite{HHH12}. These are the real AdS domain wall solutions. They are found by taking $\phi$ and $\tilde A_i$ real at the SP, and by taking $b^2$ to be real and negative for real $u$. A convenient representation of this class is provided by a contour $C_C$ that runs along the imaginary axis in the complex $\tau$-plane to an endpoint $\upsilon$ (cf Fig \ref{contours}(c)). As we will see below these saddle points do not describe classical universes and are of little interest in cosmology. Instead they are associated with Euclidean asymptotic AdS configurations. 

\section{Action of Complex Saddle Points}
\label{saddlept2}

The saddle point action is given by an integral over time along a contour in the complex $\tau$-plane connecting the SP to the endpoint $\upsilon$. The result is independent of the choice of contour. We first consider the action $I_1$ of the first class of saddle points associated with Lorentzian histories. The asymptotic contribution to the saddle point action coming from the integral along the vertical branch of the contour $C_A$ is purely imaginary. This is because the integrand is real and $d\tau=idy$. The real part $I_R$, which governs the probabilities of the de Sitter histories, tends to a constant along the $x=x_r$ while the imaginary part $S (\upsilon) \propto e^{3Hy_{\upsilon}}$. 

The AdS representation based on the contour $C_B$ provides a different way to calculate $I_R$. To see this we first consider the action integral $I_h$ along the horizontal branch of the contour in Fig \ref{contours}(b). Using the Hamiltonian constraint this can be written as,
\be
\label{horizhij}
I_h(\upsilon_a,\upsilon)=\frac{1}{8\pi}\int_{x_a}^{x_r}dx\int d^3x \ g^{\frac{1}{2}} \left[6H^2 -{^3}R + 8\pi V(\phi) +4\pi ({\vec\nabla}\phi)^2 + 4\pi F^{ij}F_{ij} \right]
\ee
where ${^3}R$ is the scalar three curvature of $g_{ij}$. The expansions \eqref{hijexpn} - \eqref{vectexp} imply that the leading contribution to $I_h$ from the vector term is ${\cal O}(u)$ and therefore negligible in the asymptotic limit. Further, the contribution to the asymptotically finite part of the action from the scalar and gravitational terms in \eqref{horizhij} vanishes as a consequence of the asymptotic Einstein equations \cite{HH11}. Hence the asymptotically constant term is the same on both ends of the horizontal branch. This means the tree level probabilities of different dS histories can be obtained from the asymptotically AdS region of the saddle points.

The action integral $I_a$ along the vertical $x=x_a$ branch in the AdS representation exhibits the usual volume divergences, but $I_h$ regulates these \cite{HH11}. Specifically we find \cite{HH11}
\be
\label{horizhij2}
I_h(\upsilon_a,\upsilon) = -S_{st}(\upsilon_a) +S_{st} (\upsilon)+{\cal O}(e^{-Hy_{\upsilon}})
\ee
where $S_{st}$ are the universal gravitational and scalar AdS counterterms, which arise here as surface contributions to the action which are kept \cite{HH11}. The second term in \eqref{horizhij2} gives a universal phase factor (since the signature at $\upsilon$ differs from that at $\upsilon_a$), and the first term regulates the divergences of $I_a$. At large $\upsilon_a$
\be
I_a(\upsilon_a) -S_{st}(\upsilon_a)  \rightarrow -I^{\rm reg}_{aAdS}
\label{reg}
\end{equation}
where $-I^{\rm reg}_{aAdS}[\th_{ij},\tilde \alpha,\tilde B_i]$ is the regulated asymptotic AdS action. It is a function of he asymptotic profiles of the geometry and matter fields in the AdS regime of the saddle points, which are locally given by the argument of the wave function at $\upsilon$ through eq. \eqref{phase}.

Thus we find that in the large volume limit, the action of a general, inhomogeneous saddle point of the NBWF associated with an asymptotically local dS history can be expressed in terms of the regulated action of a {\it complex} AdS domain wall and a sum of purely imaginary, universal surface terms, 
\be
I_1 [h,\chi,B_i ] = -I^{\rm reg}_{aAdS} [\th_{ij}(\vx),\tilde \alpha(\vx),\tilde B_i(\vx)] +S_{st} (\upsilon) +  {\cal O}(e^{-Hy_\upsilon})
\label{scnbwf}
\ee
The probabilities of the asymptotically de Sitter histories are governed by the real part of $I_1$ and hence given by
\be
\Re [I_1(\upsilon)] \equiv I_R(\upsilon) = -\mathrm{Re} [I_{aAdS}^{\rm reg} ] \ .
\label{aprobs}
\ee

It is straightforward to evaluate the action $I_2$ of the second class of real, Euclidean asymptotically AdS saddle points. This is simply given by the action integral $I_a$ along the vertical line from the SP to the endpoint $\upsilon$ in Fig \ref{contours}(c). Since all fields are everywhere real from an AdS viewpoint, the resulting action is real. It is given by
\be
I_2 [h,\chi,\tilde B_i ] = -I^{\rm reg}_{aAdS} [\th_{ij}(\vx),\tilde \alpha(\vx),\tilde B_i(\vx)] + S_{st} (\upsilon)
\label{scnbwf2}
\ee
where the boundary values $\th_{ij}(\vx)$, $\tilde B_i$ and $\tilde \alpha$ are all real. The leading surface term in $S_{st}$ is real and grows as the volume of AdS. Hence the amplitude of asymptotically AdS configurations is low relative to the classical, asymptotically de Sitter histories discussed above.

\section{Holographic Representation}
\label{holo}

The representation of {\it both} classes of saddle points in terms of Euclidean AdS `domain wall' geometries with real or complex matter profiles leads to a holographic formulation of the NBWF by applying the Euclidean version of AdS/CFT. This relates the asymptotic AdS factor $\exp (-I^{reg}_{aAdS}/\hbar)$ in the wave function to the partition function of a Euclidean dual field theory in three dimensions \cite{Witten98,Horowitz04},
\be
\exp (-I^{reg}_{aAdS}[\tilde h_{ij},\tilde \alpha,\tilde B_i]/\hbar) = Z_{QFT}[\tilde h_{ij},\tilde \alpha,\tilde B_{i}] = \langle \exp \int d^3x \sqrt{\tilde h} \left( \tilde \alpha {\cal O} +\tilde B_{i} J^{i} \right)\rangle_{QFT}
\label{operator}
\ee
The dual QFT is defined on the conformal boundary represented here by the three-metric $\th_{ij}$. The operator ${\cal O}$ is constructed from scalars in the boundary theory and $\vec J$ is a current. The brackets $\langle \cdots \rangle$ on the right hand side denote the functional integral average involving the boundary field theory action minimally coupled to the metric conformal structure represented by $\tilde h_{ij}$. For the spherical domain walls corresponding to homogeneous histories this is the round three-sphere, but in general $\tilde \alpha, \tilde B_{i}$ and $\tilde h_{ij}$ are functions of the boundary coordinates $\vx$. {\uf The three-divergence of the current  $\vec J$  vanishes ensuring the invariance of the right hand sided of \eqref{operator} under gauge transformations of $\tilde B_i$ matching the invariance of the AdS action on the left \cite{Witten98}.} 
Applying \eqref{operator} to \eqref{scnbwf} and \eqref{scnbwf2} yields
\be
\label{dscft}
\Psi (\upsilon) = \frac{1}{Z_{QFT}^{\epsilon}[\tilde h_{ij},\tilde \alpha,\tilde B_{i}] }\exp(-S_{st}(\upsilon)/\hbar)
\ee
where $\epsilon\sim 1/|Hb|$ is a UV cutoff \cite{HH11}. The surface terms are imaginary for the complex saddle points associated with Lorentzian histories, and real for the saddle points corresponding to Euclidean AdS configurations. {\uf The action $S_{\rm ct}$ is independent of the vector field. This together with the gauge invariance of $Z_{QFT}$ ensures that the wave function satisfies the $\vec\nabla\cdot\vec E=0$ constraint of Maxwell theory. }

Equation \eqref{dscft} is an example of a semiclassical dS/CFT duality. The arguments of the wave function enter as external sources in the dual partition function that turn on deformations (except for the scale factor $b=c/u$ which enters as a UV cutoff). The dependence of the partition function on the values of the external sources yields a holographic no-boundary measure on the space of asymptotic configurations. For boundary configurations given by sufficiently small values of the matter sources and sufficiently mild deformations of the round three sphere one expects the integral defining the partition function to converge, yielding a non-zero amplitude. The holographic form \eqref{dscft} of the NBWF thus involves a range of different deformations of a single underlying CFT. The sources $\tilde \alpha$ and $\tilde B_{\mu}$ of the deformations are real for saddle points associated with asymptotically AdS configurations. By contrast the sources are complex for saddle points associated with Lorentzian, classical, asymptotically de Sitter histories. In the latter regime of superspace the dual form of the NBWF thus involves complex deformations of a Euclidean CFT.

\section{Vector Fields in Homogeneous Backgrounds}
\label{homo}

As an illustration we compute the asymptotic NBWF of vector perturbations in a homogeneous and isotropic, expanding background \cite{Louko88}. Here we have in mind that the scalar field and $\Lambda$ are responsible for the background evolution. The details of this do not matter, because the vector is conformally invariant and hence decouples to quadratic order. We evaluate the semiclassical vector wave function first using the dS representation of the background saddle point and then using its AdS representation.

The line element of a homogeneous and isotropic saddle point can be written as
\begin{equation}
ds^2=N^2(\lambda) d\lambda^2 +a^2(\lambda) \gamma_{ij} dx^i dx^j
\label{eucmetric}
\end{equation} 
where $\gamma_{ij}$ is the metric on a unit round three sphere. Homogeneous and isotropic minisuperspace is spanned by the boundary value $b$ of the scale factor and the value $\chi$ of the scalar field. {\uf Neglecting the back reaction of the Maxwell field on the background}, we write
\begin{equation}
\Psi(b,\chi,\vec B)  =  \Psi_0 (b,\chi) \psi_V(b,\chi,\vec B),
\label{semiclassback}
\end{equation}
where $\Psi_0 (b,\chi)$ is a background wave function given by the action of a saddle point solution $(a(\tau),\phi(\tau))$ that matches $(b,\chi)$ at the boundary of the disk, and is regular elsewhere \cite{HHH08,HHH08b}. We assume the phase of $\Psi_0$ varies rapidly so that it predicts a classical background. The wave function $\psi_V$ of vector perturbations is defined by the remaining integral over $A_{\mu}$,
\begin{equation}  
\psi_{\uf V}(b,\chi,\vec B) \equiv  \int_{\cal C}\delta A_{\mu}  \exp(-I_V^\epb[a(\tau),\phi(\tau),A_{\mu}(\tau,\vx)]/\hbar) .
\label{qftwf}
\end{equation}
where the integral is over all regular histories on a disk which match $\vec B$ on its boundary.

We evaluate \eqref{qftwf} in the steepest descents approximation. The Euclidean action $I_V^\epb$ of vector perturbations around saddle points of the form \eqref{eucmetric} is given by
\be
I_V^\epb = \int \sqrt{\gamma}\left[ \frac{a}{2N} \left( \gamma^{ij} A_{i,\lambda}A_{j,\lambda} + A_0 (2A^k_{\ ;k0} - A_{0;k}^{\ \ \ k} \right)+ \frac{N}{2a} A^k (2A_k + A^i_{\ ;ik} - 
A_{k;i}^{\ \ \ i}) \right]
\label{vectact} 
\ee
Working in Coulomb gauge $A_0=\nabla^i A_i =0$ and in terms of conformal time $\eta$ defined as $a d\eta = N d\lambda$, the solutions take the following form,
\be
A_i (\eta,\Omega)= \sum_{nlmp} f_{nlmp} (\eta) \left( S_i^{(p)} \right)^n_{lm}
\ee
where $\left( S_i^{(p)} \right)^n_{lm}$ are the transverse eigenfunctions of the vector Laplacian $\Delta$ on the three-sphere, with $\Delta S_i = -(n^2-2)S_i$. From here onwards we denote the indices collectively by $(n)$. The solutions for $f_{(n)}$ are proportional to $e^{\pm n\eta}$. The no-boundary condition of regularity at the SP selects the solution that decays as $\eta \rightarrow -\infty$.
Hence we get
\be
\label{vecsol}
A_i (\eta,\Omega) = \sum_{(n)} B_{(n)}e^{-in\theta_{\upsilon}} e^{n\eta} S_i^{(n)} 
\ee
where $B_{(n)}$ are the mode coefficients of the boundary configuration $\vec B$ and $i \theta_{\upsilon}$ is the imaginary part of $\eta$ at the boundary surface\footnote{When the background is the empty de Sitter saddle point $\theta_{\upsilon} =\pi/2$.}. The asymptotic expansion of \eqref{vecsol} in terms of the complex time variable $u$ defined in \eqref{defu} is of the form \eqref{vectexp},
\be
A_i = \sum_{(n)} B_{(n)} \left(1 - 2ine^{-ix_r} u + {\cal O}(u^2) \right) S_i^{(n)} 
\ee
where we have used that $\eta = i\theta_{\upsilon} -2ie^{-ix_r}u + {\cal O}(u^2)$ in the large volume regime. Evaluated on solutions the vector action \eqref{vectact} reduces to a surface term,
\be
I_V^\epb = \frac{1}{2} \int d^3x (\gamma)^{1/2} \gamma^{ij} A_{i,\eta}A_{j}
\ee
Substituting \eqref{vecsol} we get
\be
I_V^\epb  = \sum_{(n)} \frac{n}{2} B^2_{(n)}
\label{vecds}
\ee
This is real and positive, indicating that the vector perturbations are in their quantum mechanical ground state.

We could equally well have computed the vector wave function using the AdS representation of the saddle point. The saddle point action of a vector field $\tilde A_{\mu}$ with boundary value $\tilde B_i$ in a homogeneous and isotropic, Euclidean, asymptotically AdS background is obtained in a similar manner and given by
\be
I^{\epb}_{V,aAdS} = \sum_{(n)} \frac{n}{2} \tilde B^2_{(n)}
 \label{adsvec}
\ee
Eq. \eqref{aprobs} implies this gives the wave function of vector perturbations in a classical, homogeneous and isotropic, asymptotically de Sitter background when $\tilde B_i$ is taken to be purely imaginary. Substituting $\tilde B_{i} = iB_{i}$ in \eqref{adsvec} and using \eqref{aprobs} indeed yields \eqref{vecds}.

\section{Discussion}
\label{disc}

We have generalized the holographic formulation of the NBWF to models including vector matter. We first derived an `AdS representation' of the NBWF saddle points associated with classical, asymptotically dS universes with scalar and vector matter. In this representation, the saddle point geometry consists of a Euclidean AdS domain wall which joins smoothly onto an asymptotically real, Lorentzian, de Sitter geometry through a transition region where the spatial metric changes signature. The scalar and vector matter profiles along the AdS domain wall are complex. The relative probabilities of different classical boundary configurations $(h,\chi,\vec B)$ are given by the regularized AdS domain wall action.

Applying Euclidean AdS/CFT then yields a holographic form of the semiclassical NBWF in terms of the partition function of a dual, Euclidean CFT with complex sources specified by the asymptotic scalar and vector boundary values in the AdS regime of the saddle points.  The dual description of the no-boundary measure on classical, Lorentzian configurations thus involves a generalization of Euclidean AdS/CFT to CFTs with complex deformations. The phases of the sources are independent of the specific quantum state and of the dynamics in the interior. They are fully specified by the requirement that the fields be real at the final de Sitter boundary. The phase of the scalar source in the dual is $e^{-i\lambda_{-} \pi/2}$, where $\lambda_{-}= \frac{3}{2}[1-\sqrt{1-(2m/3)^2}]$, and the vector field source is purely imaginary\footnote{\uf Mithani and Vilenkin \cite{Vilenkin13} have considered a scheme in which the vector field is real from an AdS viewpoint. This does not give a well-defined theory (as they point out).}.

From a holographic viewpoint it is natural to consider the wave function on an extended domain that includes real deformations of the dual CFT \cite{HHH12,HHH12b}. Partition functions with real deformations specify the amplitude of real, asymptotically AdS boundary configurations. In the bulk their amplitude is given by real Euclidean AdS domain walls. These are of little relevance in cosmology, since the lack of a rapidly varying phase factor in the action means they are not associated with classical histories. In fact the AdS volume contribution to the action heavily suppresses their contribution to the NBWF. Nevertheless, the inclusion of both sets of saddle points in the configuration space - as suggested by holography - yields an appealing, signature-invariant formulation of the wave function defined on an extended configuration space with three-metrics of both signatures \cite{HHH12}. One signature corresponds to deformations with real sources and the other to complex sources.

From a bulk viewpoint the relation between (asymptotic) Euclidean AdS and Lorentzian de Sitter exhibited here is reminiscent of a `symmetry' of the action \eqref{curvact}-\eqref{mattact} under a reversal of the signature of the metric \cite{Vilenkin88}. A signature reversal relates the action of a dS theory of gravity coupled to a scalar with a positive potential $V$ and a vector field $A_{\mu}$ to the action of an AdS theory of gravity coupled to a scalar with potential $-V$ and a vector field $\tilde A_{\mu} = iA_{\mu}$ that is imaginary from a dS viewpoint. This is not unlike the transition from supersymmetry in AdS to pseudo-supersymmetry in dS, where the vectors in the de Sitter theory exhibit a similar behavior. In that context as well it has been argued that the dS and AdS theories should really be viewed as two real domains of a single underlying complexified theory \cite{Bergshoeff07,Skenderis07}. The complex structure of the wave function in our scheme provides a natural setup for a unified description of this kind. This is made explicit in its holographic form which involves complex deformations of a single underlying dual conformal field theory.

Our analysis so far applies only to toy models of de Sitter gravity or, in AdS terms, to consistent truncations of AdS supergravity theories that retain only low mass scalars, and vectors. A full realization of holographic cosmology in string theory remains an open problem. A particularly intriguing issue has to do with the tower of irrelevant operators featuring in the usual AdS/CFT duals. These correspond to tachyonic fields in the de Sitter domain of the theory. The requirement of an asymptotic de Sitter structure in dS/CFT amounts to a final boundary condition on these fields.
More generally the asymptotic dS structure implies that the asymptotic geometry and fields behave classically \cite{HHH12}. The no-boundary condition of regularity on the saddle points implements the final boundary condition on tachyons since it excludes classical histories in which these are not set to zero \cite{HHH08,HM13}. 

The dual partition function is not concerned with the classical evolution itself - which is governed by the phase factor of the wave function \eqref{semiclass} - but merely calculates the probability measure on classical phase space in the no-boundary state \cite{Horowitz04,HH11}. Given that a probabilistic interpretation of the wave function of the universe is restricted to its classical domain anyway \cite{HHH08} this suggests dS/CFT should be viewed as a statement about the coarse-grained, classical predictions of the wave function only\footnote{This is also born out by calculations of the wave function of perturbations in dS/CFT where the dual partition function captures the wave function of super-horizon modes which behave classically \cite{Maldacena03,HH11}.}. But it would be very interesting to understand the origin and nature of the late-time classicality constraint from a dual viewpoint.

\vskip .2in

\noindent{\bf Acknowledgments:} We thank {\uf Gary Horowitz, Audrey Mithani, Joe Polchinski,  Steve Shenker, Neil Turok and Alex Vilenkin} for helpful discussions. This research was supported in part by the US National Science Foundation Grant No. NSF PHY11-25915 and No. NSF PHY08-55415, and by the National Science Foundation of Belgium under the FWO-Odysseus program.

\end{document}